\documentclass[a4paper,11pt]{article}
\usepackage[margin=1in]{geometry}
\usepackage[utf8]{inputenc}
\usepackage{authblk}
\usepackage{graphicx, xcolor}
\usepackage{amsmath}
\usepackage{amsfonts}
\usepackage{amssymb}
\usepackage{fancyhdr}

\usepackage{subcaption}
\usepackage[font=small,labelfont=bf]{caption}

\usepackage[hidelinks]{hyperref}  

\newcommand{\lambdab}{\bar{\lambda}}
\newcommand{\mub}{\bar{\mu}}

\def\lesssim{\mathrel{\spose{\lower 3pt\hbox{$\mathchar"218$}}
 \raise 2.0pt\hbox{$\mathchar"13C$}}}
\def\gtrsim{\mathrel{\spose{\lower 3pt\hbox{$\mathchar"218$}}
 \raise 2.0pt\hbox{$\mathchar"13E$}}}

\title{{\bf essHi-C: Essential component analysis of Hi-C matrices}}

\author[1]{Stefano Franzini}
\author[2]{Marco Di Stefano}
\author[1,*]{Cristian Micheletti}

\affil[1]{\footnotesize SISSA, International School for Advanced Studies, Trieste, I-34136, Italy.}
\affil[2]{\footnotesize CNAG-CRG, Centre Nacional d’An\'alisi Gen\'omica - Centre de Regulaci\'o Gen\'omica, Barcelona, 08028, Spain.}
\affil[*]{michelet@sissa.it}
\date{}

\begin{document}

\maketitle


\abstract{
\noindent \textbf{Motivation:} Hi-C matrices are cornerstones for qualitative and quantitative studies of genome folding, from its territorial organization to compartments and topological domains. The high dynamic range of genomic distances probed in Hi-C assays reflects in an inherent stochastic background of the interactions matrices, which inevitably convolve the features of interest with largely aspecific ones.\\
\textbf{Results:} Here we introduce a discuss essHi-C, a method to isolate the specific, or essential component of Hi-C matrices from the aspecific portion of the spectrum that is compatible with random matrices.
Systematic comparisons show that essHi-C improves the clarity of the interaction patterns, enhances the robustness against sequencing depth, allows the unsupervised clustering of experiments in different cell lines and recovers the cell-cycle phasing of single-cells based on Hi-C data. Thus, essHi-C provides means for isolating significant biological and physical features from Hi-C matrices.\\
}

\section*{Introduction}
Much of our current understanding of the structural-functional interplay in the genome owes to the advancements fostered by chromosome conformation capture-based methods (3C \cite{Dekker_2002}), which are now numerous \cite{Grob_2018,Ubelmesser_2019}.
For instance, Hi-C experiments demonstrated that inter-chromosome interactions are suppressed compared to intra-chromosome ones, thus giving quantitative support to the earlier notion of chromosome territories \cite{Cremer_2001}. Inspection of the interaction matrices, and their dominant eigenvector \cite{Lieberman_Aiden_2009}, revealed the existence of two main chromatin compartments. Analysis of the interaction patterns showed the presence of self-interacting domains termed topologically associating domains (TADs) \cite{Sexton_2012,Dixon_2012,Nora_2012}, which may form complex nested structures \cite{Fraser_2015}.

The increasing Hi-C resolution has made it possible to compare interaction patterns of different samples. The long term goal of such comparative studies is establishing which aspects of genome organisation are varied across different stages of cell development \cite{Zheng_2019} and cell fate \cite{Bonev_2017,Stadhouders_2018,Paulsen_2019,Sati_2020}, are affected by differences in gene transcription \cite{Dekker_2013,Sexton_2015,Schmitt_2016}, or are mis-regulated in disease-related phenotypic alterations \cite{Lupianez_2015} and cancer \cite{Krijger_2016}.

Owing to the importance of comparative analysis, it is increasingly crucial to cross-validate data gathered with different protocols, resolution, and sequencing depth \cite{Fortin_2015,Yang_2017,Stansfield_2018,Yardimci_2019,Zhou_2019}, thus identifying the common and statistically-significant features \cite{Ursu_2018} in Hi-C matrices.

These observations pose, in turn, the more fundamental question of whether it is at all feasible to identify {\em a priori} the robust, significant features of a given Hi-C matrix without the necessity to resort to other terms of comparisons, which might contain biases or even not be available at all. An {\em a priori} knowledge of the significant features of a Hi-C matrix would also enhance the capability of detecting meaningful differences and similarities with other Hi-C matrices.

In this study, we show that spectral analysis methods, which rely on the information content of eigenvectors and eigenvalues, are ideally suited to this endeavour. However, with only a few exceptions, spectral methods have so far focussed on the first one or two eigenvectors of Hi-C matrices, which are informative for the chromatin compartmentalisation.

Here, we introduce essHi-C (after 'essential Hi-C') to extend these considerations to the full spectrum of Hi-C matrices, which we regularise for genomic distance. We show that most of the spectrum is compatible with that of random matrices, and thus represents an aspecific component shared across chromosomes from different samples. Interestingly, by discounting this aspecific part of the spectrum, and retaining only what we term the essential component, we enhance the definition of chromosomes' architectural features and provide a significant advantange to readily pick up similarities of replicates and dissimilarity across different cell lines. Accordingly, heterogeneous sets of Hi-C matrices can be reliably grouped per cell lines using unsupervised clustering, which can even pick up the elusive interaction signatures of distinct experimental protocols. Finally, we show that essential matrices are stable against variations of the sequencing depth and the amount of input material of Hi-C. We found that essHi-C is predictive of the features discernible in Hi-C matrices with deeper sequencing and is informative for the genome architecture in extremely low-input Hi-C datasets, like the ones from single-cell Hi-C experiments.


\section*{Methods}

{\bf Dataset.} Our dataset consist of intra-chromosome Hi-C matrices from $79$ experiments of $9$ human cell lines, including five with normal karyotype (GM12878, IMR90, NHEK, HMEC, hESC) and four with cancerous karyotypes (T47D, K562, KBM7, SKBR3) (see Supplementary Table 1). The sra-toolkit (\href{http://ncbi.github.io/sra-tools/}{http://ncbi.github.io/sra-tools/}) (v2.9.6) was used to fetch the Hi-C datasets from the public sequence read archive (SRA) (Supplementary Table 1) and convert them to FASTQ format after validation.  The TADbit pipeline \cite{Serra_2017} (\href{https://github.com/3DGenomes/tadbit}{https://github.com/3DGenomes/tadbit}) was used to (i) check the quality of the FASTQ files; (ii) map the paired-end reads to the {\em H. sapiens} reference genome (release GRCh38/hg38) using GEM \cite{Marco_2015} accounting for restriction enzyme cut-sites; (iii) remove non-informative reads using the default TADbit filtering options; (iv) merge datasets within each experiment when appropriate; (v) normalise each experiment using the OneD method \cite{Vidal_2018} at 100 kbp (kilo-basepairs) resolution.

{\bf Observed over expected normalisation.}
The obtained intra-chromosomal Hi-C matrices were subject to the so-called observed over expected (OoE) normalisation \cite{Lieberman_Aiden_2009} to discount the overall dependence of matrix entries, $A_{ij}$, on the corresponding genomic distance, ${s=|i-j|}$. Starting from a OneD-normalised matrix, $A$, the average interactions at any genomic distance $s$ was computed, $I(s)$. Each entry of the normalised matrix, $B$, is then defined as $B_{ij} = \frac{A_{ij}}{I(s=|i-j|)}$.\\
Except when otherwise stated all Hi-C matrices considered in their full form are intended to be OoE-normalised.

{\bf Random matrix ensemble.} As a null model for the OoE Hi-C matrices we considered the so-called Gaussian orthogonal ensemble, whose elements are symmetric matrices with entries drawn from a unitary Gaussian distribution, hence with zero mean and unit variance.
On average, the salient spectral properties of the elements of this ensemble are as follows \cite{O_Rourke_2016,Livan_2018}. The set of orthonormal eigenvectors sample uniformly the surface of a unit ($N-1$)-sphere, where $N$ is the linear size of the matrices. The generic component, $x$, of any eigenvector follows the same Gaussian distribution with zero mean and variance equal to $1/N$,
\begin{equation} \label{Eq:eigen_components}
  p(x) = \sqrt{\tfrac{N}{2 \pi}}e^{ -\tfrac{N x^2}{2} }\ .
\end{equation}
The distribution of the eigenvalues, $\lambda$'s, is governed by the Wigner's semicircular law:
\begin{equation} \label{Eq:wigner}
  p(\lambda) = \frac{1}{2 \pi \Lambda^2} \sqrt{ 4 \Lambda^2 - \lambda^2}
\end{equation}
\noindent with the interval $[-\Lambda, \Lambda]$ being the support of $\lambda$.

\begin{figure}[!tpb]
\centerline{\includegraphics[width=1.\linewidth]{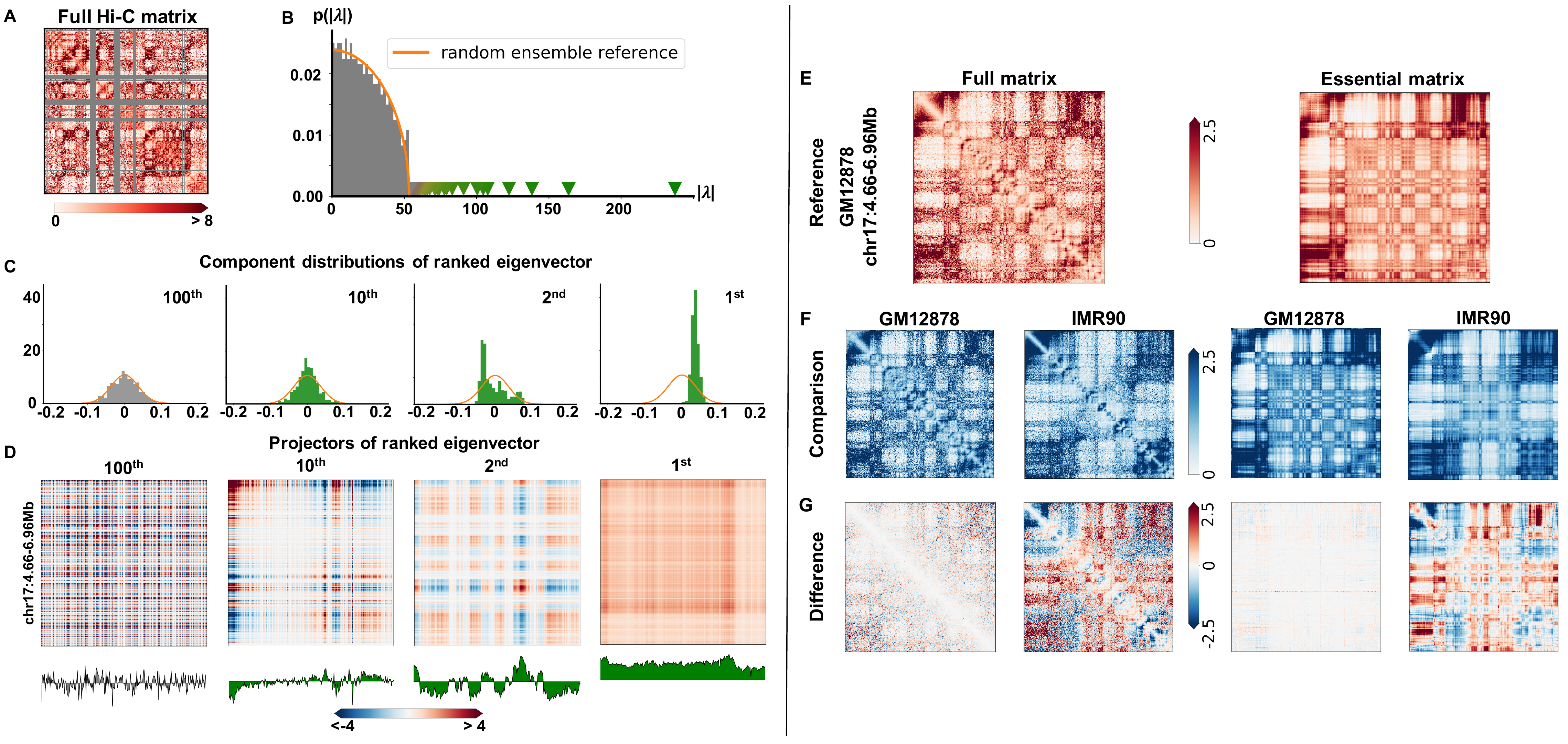}}
\caption{{\bf Spectral properties of Hi-C vs random matrices and comparison of full and essential Hi-C matrices.}
{\bf A} Hi-C matrix of human chromosome 17 of cell line GM12878
(experiment HIC001) after OoE normalisation, see Methods. Grey bands
correspond to the centromere and other regions removed by the TADbit
filtering step (see Methods). Its eigenvalue distribution is shown in
panel {\bf B} along with the distribution for random unitary random
matrices of same size (yellow). The outlier part of the Hi-C matrix
spectrum, $| \lambda | > 58.7$, is highlighted in green. {\bf C}
Probability distributions for the components of Hi-C matrix
eigenvectors of rank 1, 2, 10 and 100; analogous distribution for
random matrix eigenvectors is shown in yellow. The spectral projectors
of the Hi-C eigenvectors are shown in panel {\bf D} for the region of
chr17 between 4.66 and 9.69Mb. Below them the amplitude of the
eigenvectors' components is shown. {\bf E} Full Hi-C matrix of
chromosome 17 from cell line GM12878 (experiment HIC001) and its
essential version the genomic region chr17:4.66-6.96Mb is shown for
clarity. {\bf F} Other instances
of full and essential matrices for the same genomic region for the same cell
line (GM12878 from experiment HIC002) and a different ones (IMR90 from the
experiment HIC050), see also Supplementary Table 1. The differences of these
matrices with the reference ones of panel {\bf E} are shown in panel {\bf G}.}
\label{Fig1}
\end{figure}

{\bf Essential component of Hi-C matrices.}
One of our main results is that OoE Hi-C matrices have a spectrum largely consistent with that of random matrices, except for a limited set of eigenvectors with atypically large eigenvalues in modulus. Borrowing a terminology introduced in other contexts \cite{Amadei_1993,Micheletti_2013}, we refer to these eigenspaces as {\em essential} and we use their spectral summation to define the essential Hi-C matrix. Starting from an OoE-normalised matrix, $A$, the entries of its essential form are defined as
\begin{equation}
  A_{ij}^{ess} = \sum_{n=1}^{n^*} \lambda_{n} a^{(i)}_{n} a^{(j)}_{n},
  \label{eqn:ess}
\end{equation}
\noindent where $a_{n}^{(i)}$ is the $i-$th component of the $n-$th eigenvector of $A$ and $\lambda_n$ is the associated eigenvalue. The matrix ($a^{(i)}_{n} a^{(j)}_{n}$) denotes the projector associated to the eigenvector $a$ (see Fig. \ref{Fig1}D for examples). Once weighed for the correspondent eigenvalue ($\lambda_{n} a^{(i)}_{n} a^{(j)}_{n}$), it can be interpreted as the eigenspace, that is the contribution of $a$ to the Hi-C contact pattern. The eigenspaces are ranked for decreasing modulus of the eigenvalues so that the summation is restricted to the top $n^*$ essential spaces. In principle, $n^*$ could be assigned differently for each matrix. For simplicity we set $n^*=10$ for all the applications in this study, except for the case of sparse single-cell Hi-C matrices, as the results do not vary appreciably upon including additional spaces, see Supplementary Fig.~S1.

{\bf Measuring matrix similarity across replicates.}
In our dataset, the most numerous (34) independent Hi-C experiments on the same cell line (replicates) pertains to GM12878 cells. The similarity of these {\em a priori} equivalent Hi-C measurements was assessed as follows. For each chromosome (including autosomes and X) $100$ pairs of replicate Hi-C matrices were randomly picked. For any such pair, $P$ and $Q$, we next obtained the sum and difference matrices, $S=P+Q$ and $D=P-Q$. The similarity parameter $\gamma$ was computed as
\begin{equation}
  \gamma = \frac{ \langle S \rangle }{ 2 \sigma_{D}}
\end{equation}
\noindent where the angular brackets denote the average over the all entries $S_{ij}$ with $i \ge j$ and $\sigma_D$ is the root mean-square value of all entries $D_{ij}$, again with with $i \ge j$. OoE-normalised Hi-C matrices with good (poor) similarity thus yield values of $\gamma$ that are much larger (smaller) than 1.

{\bf Measuring matrix robustness across different sequencing depths.} We considered the set of GM12878 matrices from the experiment HIC003 \cite{Lieberman_Aiden_2009}, which have the highest available sequencing depth (see Supplementary Table 1), and used them as gold standard for comparisons with matrices at much lower depths. The consistency of a generic matrix with the gold standard (same chromosome) was measured in terms of the Spearman correlation coefficient of corresponding entries.

{\bf Metric distance of essential Hi-C matrices.}
For clustering purposes, the squared distance of two full OoE-normalised matrices, $A$ and $B$, was computed as the Eucledian distance of corresponding entries,
\begin{equation}
d^2(A,B)=\sum^\prime_{i,j} | A_{ij} - B_{ij}|^2  \ ,
    \label{eqn:dplain}
\end{equation}
\noindent where the prime denotes that the sum is taken over $ i\ge j$.
The squared distance of the matrices in the essential form is instead defined as:
\begin{equation}
    d^2(A_{\rm ess},B_{\rm ess})= \sum^{n<n^*}( \lambdab_{n}^2 + \mub_{n}^2) -
    2\sum^{n,m<n^*} \lambdab_n \mub_m ( \textbf{a}_{n} \cdot \textbf{b}_{m} )^2 ,
    \label{eqn:dess}
\end{equation}
\noindent where $\textbf{a}$ and $\textbf{b}$ are the eigenvectors of $A$ and $B$, respectively, $\lambda$ and $\mu$ are the associated eigenvalues, and $n^*$ is the spectral cutoff of eq.~\ref{eqn:ess} for the essential component. The scaling factors $\lambdab$ and $\mub$ make the distance robust respect to $n^*$, see Supplementary Fig.~S1,  and are defined as
\begin{equation}
    \lambdab_n = \frac{\lambda_n}{\sum^{n<n^*} |\lambda_n | }, \ \     \mub_n = \frac{\mu_n}{\sum^{n<n^*} |\mu_n | }.
\end{equation}

{\bf Genome-wide distance of Hi-C matrices.}
The genome-wide distance of all chromosomal matrices of two Hi-C experiments, $\alpha$ and $\beta$, is defined as:
\begin{equation}\label{dist}
    d(\alpha,\beta) = \left[ \sum_{k} d^2(M_{\alpha k},M_{\beta k}) \right]^{1/2} ,
\end{equation}
where index $k$ runs over chromosomes, and $M_{\alpha,k}$ is the Hi-C matrix of chromosome $k$ from experiment $\alpha$ and, depending on the context, $d$ is either the plain Eucledian distance of eq.~\ref{eqn:dplain} or the essential one of eq.~\ref{eqn:dess}.

{\bf Single cell Hi-C analysis.} The single-cell Hi-C (scHi-C) matrices
in the haploid mouse embryonic stem cells (mESC) dataset of \cite{Nagano_2017} was considered and analysed using the TADbit pipeline as for bulk Hi-C matrices with the exception of the OneD normalisation, which is not tailored for scHi-C matrices. After discarding assays
with missing data for one or more chromosomes, our dataset consisted of $320$ complete single cell assays covering three cellular stages labelled as G1, early-S, late-S/G2 in ref.~\cite{Nagano_2017}, see Supplementary Table 2.
Stages labelled as post-M and pre-M were not included because they contained zero and only one
complete assay, respectively.

scHi-C matrices require suitable {\em ad hoc} extension of the essHi-C analysis used for standard (bulk) Hi-C matrices for their sparsity, which prevents taking meaningful OoE normalisations.
For this preliminary application we identified the $n^*=50$ top ranking eigenspaces as the essential ones because they typically suffice to capture most of the trace of the non-normalised Hi-C matrices, see Supplementary Fig.~S2. The same distance definition of eq.~\ref{eqn:dess} was used for comparing essential scHi-C matrices and compute the ROC curves, where the G1, early-S, late-S/G2 labelling of \cite{Nagano_2017} was used as gold standard.

The tripartite clustering of the essHiC dataset was computed from the Ward dendrogram.

\section*{Results}

\subsection*{Spectral comparison of Hi-C and random matrices}

Throughout the study we analysed Hi-C interaction matrices after the
observed-over-expected (OoE) normalisation. As it is shown in the instance of Fig.~\ref{Fig1}A, the OoE normalisation, which discounts genomic-distance biases \cite{Lieberman_Aiden_2009}, puts on equal footing interactions at different sequence separations. The matrix in Fig.~\ref{Fig1}A pertains to chromosome 17 of cell line GM12878 and its spectral properties are illustrated in panels B-E. Specifically, panel B shows the probability distribution of the modulus of the eigenvalues, $p(|\lambda|)$ while the
probability distributions of the components of different ranking eigenvectors are shown in panel C.
The same panels show, by contrast, the analogous quantities but computed for symmetric random matrices with the same linear size.

The two eigenvalues distributions are very closely matching throughout the range $|\lambda| \le 58.7$, which covers $95\%$ of the Hi-C spectrum. Within the $|\lambda| \le 58.7$ region, the well-matching spectral properties of Hi-C and random matrices also extends to eigenvectors. In fact, Their components are normally distributed, as for Gaussian random matrices, see panel C.

The outlying eigenvalues,  $|\lambda| > 58.7$, are highlighted in green in panel B and accounts for only a small fraction of the Hi-C matrix spectrum. The components of the associated highest-ranking eigenvectors have a manifest non-Gaussian distribution.
Thus, the top ranking eigenvalues and eigenvectors are the sole having markedly distincting properties from those found in random matrices.
This fact holds in general, as it applies to different chromosomes and cell lines, see Supplementary Figs.~S3 and S4, and thus establishes that the bulk of Hi-C matrix spectrum is largely compatible with that of random matrices, except for the small set of outlier eigenvalues and associated eigenvectors.

\subsection*{Essential Hi-C matrices}

Because the bulk of the spectrum of Hi-C matrices can be described by a statistical model informed solely by the linear size of the matrix, we discounted this aspecific component from the matrices so to isolate their {\em essential} component.  The latter, that we term essential Hi-C matrix or essHi-C for brevity, is obtained from the spectral summation of the $n^*=10$ highest ranking projectors, see Methods and Fig.~\ref{Fig1}D.

A comparison of a full Hi-C matrix and its essential component is provided in Fig.~\ref{Fig1}E. The data are for the same entry of Fig.~\ref{Fig1}A, chromosome 17 and cell line GM12878, but are presented for a chromosome portion only to aid visualization.
Remarkably, the essential matrix not only retains the distinctive contact patterns, but it presents them with greater clarity and contrast. The intra- and inter-domain contact patterns, as well as the domain boundaries, are more clearly defined too.

\subsection*{Enhancement of specificity}

The benefits of resorting to essential matrices, thus discounting the aspecific spectral component, further emerges from the comparative analysis of Fig.~\ref{Fig1}F, where the full and essential Hi-C matrices of panel A are compared with two other instances of chromosome 17, one from a biological replicate of the same GM12878 cell line, and one from the different IMR90 cell line.

The entry-by-entry subtraction of the matrices is presented in Fig.~\ref{Fig1}G and shows that contact pattern differences are sharper and more deeply marked for the essential matrices. Importantly, the difference matrix of GM12878 replicates presents as a uniform background while speckled patterns are clearly discernible for the full matrices of the different cell lines. The result holds generally and is not dependent on the considered cell types, see Supplementary Fig.~S5.

\begin{figure}[!tpb]
\centerline{\includegraphics[width=0.6\linewidth]{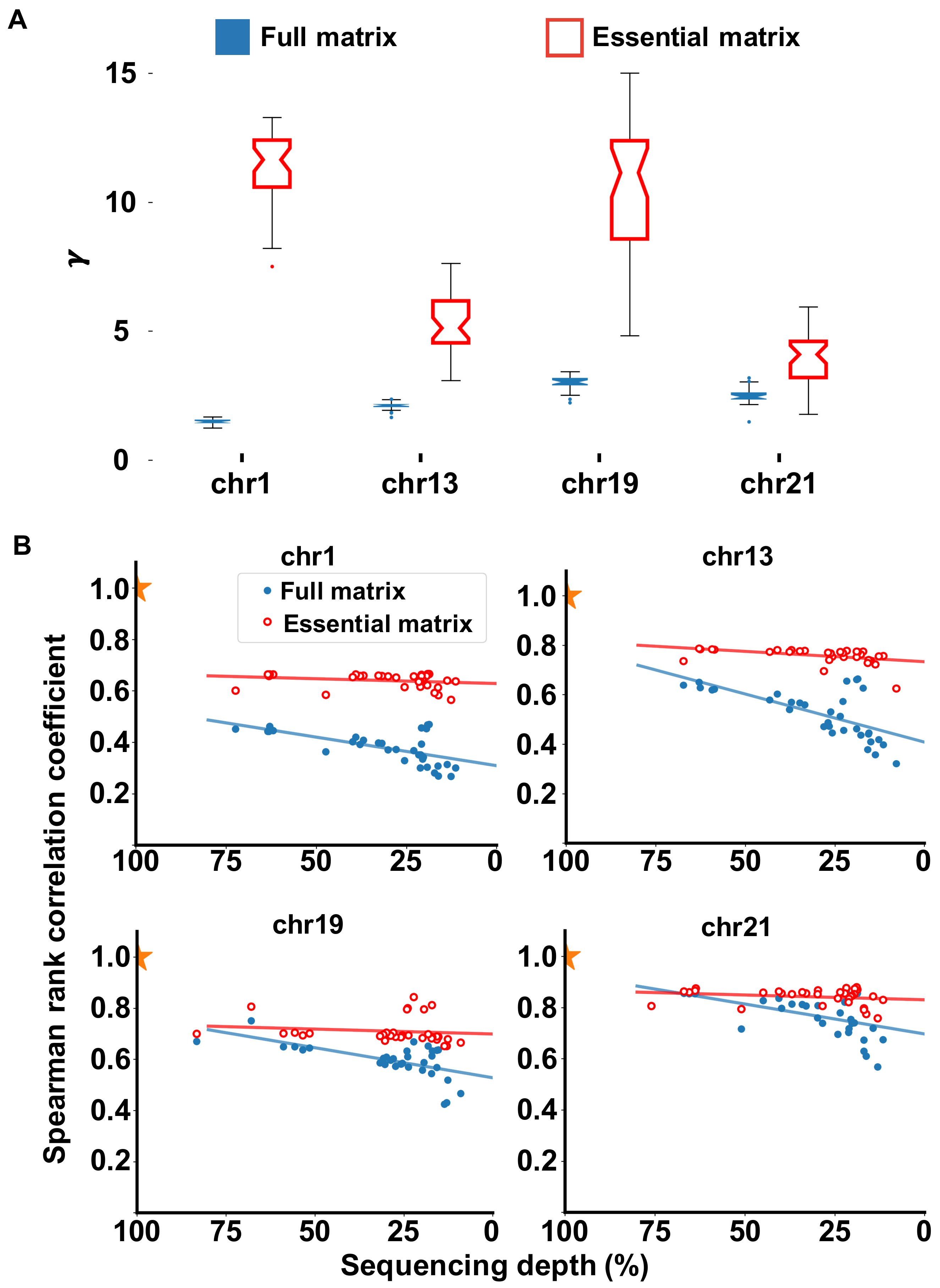}}
\caption{{\bf Matrix robustness across replicates and sequencing depths}. {\bf
A} Box-whisker plots of the consistency parameter $\gamma$, see
Methods, measured for distinct pairs of full and esential matrices for
the GM12878 cell lines. Four chromosomes with large difference of
length and gene cdensity are considered. Boxplots show: central line,
median; box limits, 75th and 25th percentiles; whiskers, 1.5 times the
interquartile range; outliers beyond this range are shown as individual points. {\bf B} Spearman rank
correlation coefficient measured for corresponding entries of a high
sequencing depth matrix (gold standard HIC003, 100\% depth) and
matrices at lower depths. Data are for the GM12878 and the same four
chromosomes as in panel {\bf A}.}
  \label{Fig2}
\end{figure}

\subsection*{Robustness across replicates and varying sequencing depths}

The enhancement of the contact pattern specificity is conveyed in Fig.~\ref{Fig2}A in terms of the similarity parameter $\gamma$. The latter quantifies how similar are corresponding entries in two matrices relative to their inherent statistical uncertainty, see Methods. Fig.~\ref{Fig2}A shows the distribution of $\gamma$ computed for randomly picked pairs of GM12878 matrices for four chromosomes. These correspond to chromosomes 1, 13, 19 and 21, which were chosen for their diverse gene content and length.

The typical values of $\gamma$ for full matrices are of the order unity and have a mild increasing dependence on chromosome length. Using, instead, the essHi-C version yields dramatic boost of the similarity parameter by about one order of magnitude, and with no particular bias or dependence on chromosome length (Fig.~\ref{Fig2}A).

We next examined the impact of the sequencing depth, taking as gold standard the full Hi-C matrices with the largest sequencing depth in the GM12878 dataset, see Supplementary Table 1. The Spearman's rank correlation coefficient of corresponding entries of the gold standard and other matrices with lower depths are presented in Fig.~\ref{Fig2}B, again, for chromosomes 1, 13, 19 and 21.

Even though the gold standard is constituted by full Hi-C matrices the highest correlation across all four chromosomes and all lower depths is observed for the {\em essential} matrices, not the full ones. The result is general in that it holds for other chromosomes too, see Supplementary Fig.~S6.

In addition, the correlation coefficient of the essential matrices has a visibly milder dependence on sequencing, the slope of the interpolating line being much smaller than the full case.

\begin{figure}[!tpb]
\centerline{\includegraphics[width=0.6\linewidth]{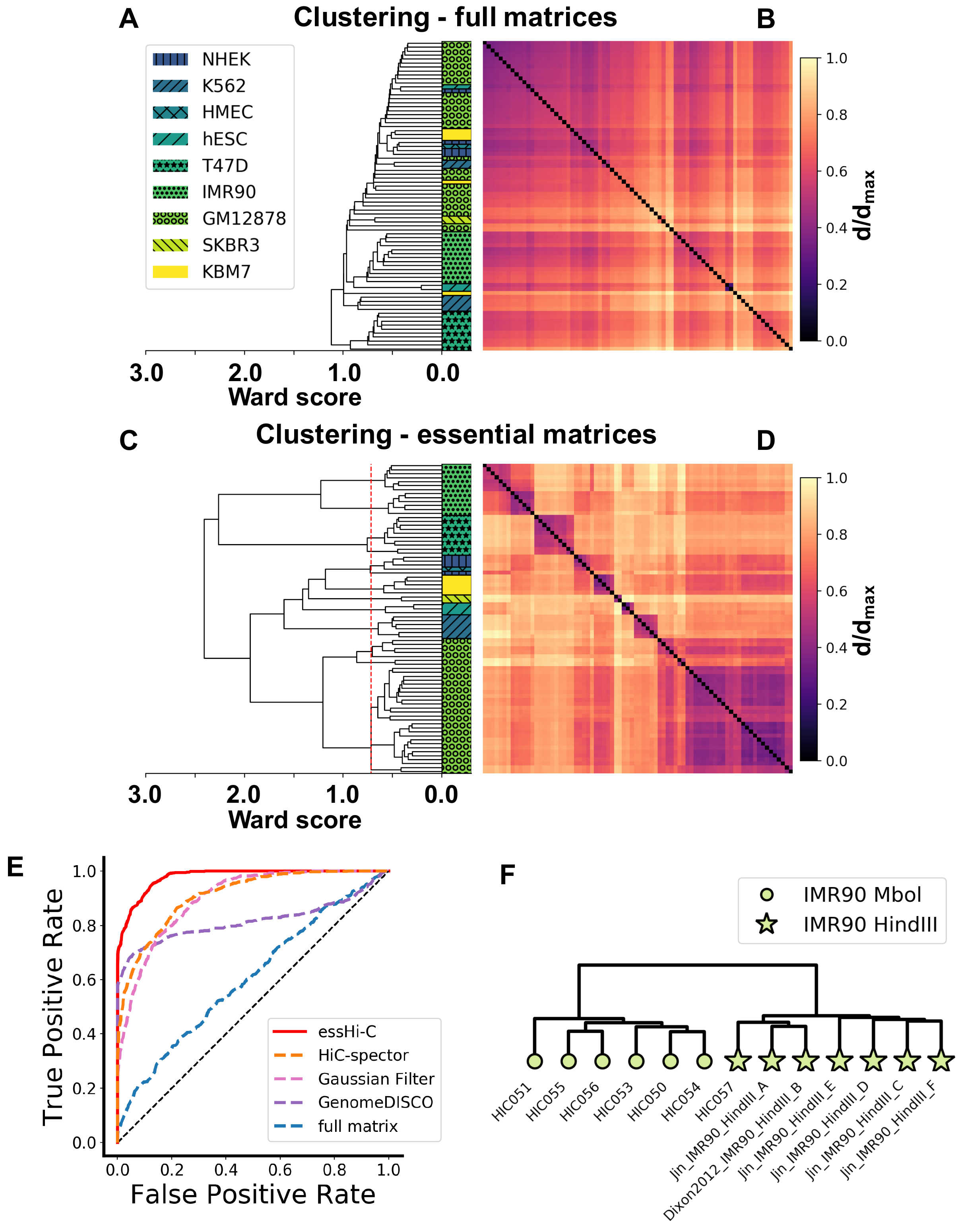}}
\caption{{\bf Clustering of different cell lines} {\bf A} Ward dendrogram of the dataset of 79 Hi-C experiments, covering different cell lines, based on the genome-wide distance of full Hi-C matrices. The pairwise distance matrix, with entries normalised to the maximum is shown in panel {\bf B}. Analogous quantities, but computed for the corresponding essential Hi-C matrices are shown in panels {\bf C} and {\bf D}. The dashed line in the dendrogram of panel {\bf C} marks the Ward score of the optimal (Dunn) subdivision in 13 clusters. The distances computed for the full and essential matrices, and other methods, were used to compute the ROC curves of panel {\bf E}, where true positives correspond to instances of the same cell lines. {\bf F} Section of the dendrogram in {\bf D} regarding the IMR90 cell line, which shows the correlation with the used restriction enzymes.}
  \label{Fig3}
\end{figure}

\subsection*{Unsupervised clustering of different cell lines}

As an application of the essHi-C analysis, we performed the unsupervised clustering of a heterogeneous dataset of $79$ matrices, comprising (non-uniformly) 8 distinct cell lines and 3 different Hi-C techniques (Supplementary Table 1 and Methods).

The two clusterings obtained by using the full and essential matrices are shown as dendograms in Fig.~\ref{Fig3}A,C. They present major and striking differences.

First, although the dendrograms are drawn using the same unit for the Ward score, the length of the branches for the essential matrices are more than twice longer ($2.5$ max Ward score) than the ones for full ones ($1.0$ max Ward score). This fact indicates that essHi-C matrices for more definite clusters compared to the full matrices.

The content of the clusters are also very different. In full matrices one can resolve the breast cancer (T47D) cell lines, whose chromosomal aberrations reflect in large-scale matrix changes \cite{Rondon_2014}. However, other cell lines are poorly resolved, including the most numerous ones of GM12878 and IMR90, Fig.~\ref{Fig3}A.

The essential matrices return instead sharp subdivisions between different cell lines already from the early stages in the dendrogram hierarchy, see panel C.  In addition, the associated pairwise distance matrix has a clear block structure, unlike the case of full matrices, see Fig.~\ref{Fig3}B,D.

The higher discriminatory power of essHi-C matrices is conveyed by the ROC curves shown in Fig.~\ref{Fig3}E.  Full matrices yield an area-under-the-curve (AUC) paramter of $0.6$, which only marginally improves with respect to the random reference ($AUC=0.5$). On the other hand essential matrices yield a nearly optimal performance, with $AUC=0.98$.

As further terms of reference, Fig.~\ref{Fig3}E shows ROC curves obtained applying other methods to the full matrices. These include a Gaussian filter (convolution with unitary or single-bin variance), GenomeDISCO \cite{Ursu_2018}, and Hi-C Spector \cite{Yan_2017}.  The Gaussian filter is often used as a means to curb noise in Hi-C matrices by averaging over neighbouring entries while the latter two methods are of interest here for they are based on the use of eigenspaces to compare Hi-C matrices. The AUC values for the Gaussian filter, Hi-C Spector and GenomeDisco are $0.90$, $0.82$ and $0.91$, respectively, which are all significant. Some of these methods, including spectral ones, were purposedly devised towards the comparative analysis of Hi-C matrices. The about optimal essHi-C performance is thus appealing as it is natively formulated as an enhancement method of individual matrices which can be adopted in comparative contexts too, as shown here.

By analyzing the quality of different numbers of groupings within the essHIC dendrogram, using the Dunn index, one discovers that the optimal number of clusters is $13$, which is larger than the number of cell lines ($9$).

All clusters except one contain one cell-type only, however, while all experiments of cell-types K562, hESC, SKBR3, and KBM7 are contained within a single cluster, other cell-types ensembles display a richer internal structure: this is apparent especially for IMR90 experiments, which are sharply divided into two clusters (Fig.~\ref{Fig3}F). Upon closer inspection one finds that this division is not an arbitrary one, but corresponds to different restriction enzymes and experimental techniques: one cluster contains {\em In-situ} experiments using Mbol, while the other are {\em dilution} Hi-C experiments using HindIII.

Other subdivisions within cell-lines are not as clear-cut and cannot be explained only in terms of different experimental methodologies. However one interesting case is given by the only mixed cluster, containing NHEK and HMEC experiments: both cell lines are epithelial samples, which may explain their similarity, moreover all experiments within the cluster share the same methodology ({\em In-situ}, Mbol). On the other hand the single isolated NHEK experiment is a {\em dilution} Hi-C, and uses the HindIII reduction enzyme.

\begin{figure}[!tpb]
\centerline{\includegraphics[width=0.6\linewidth]{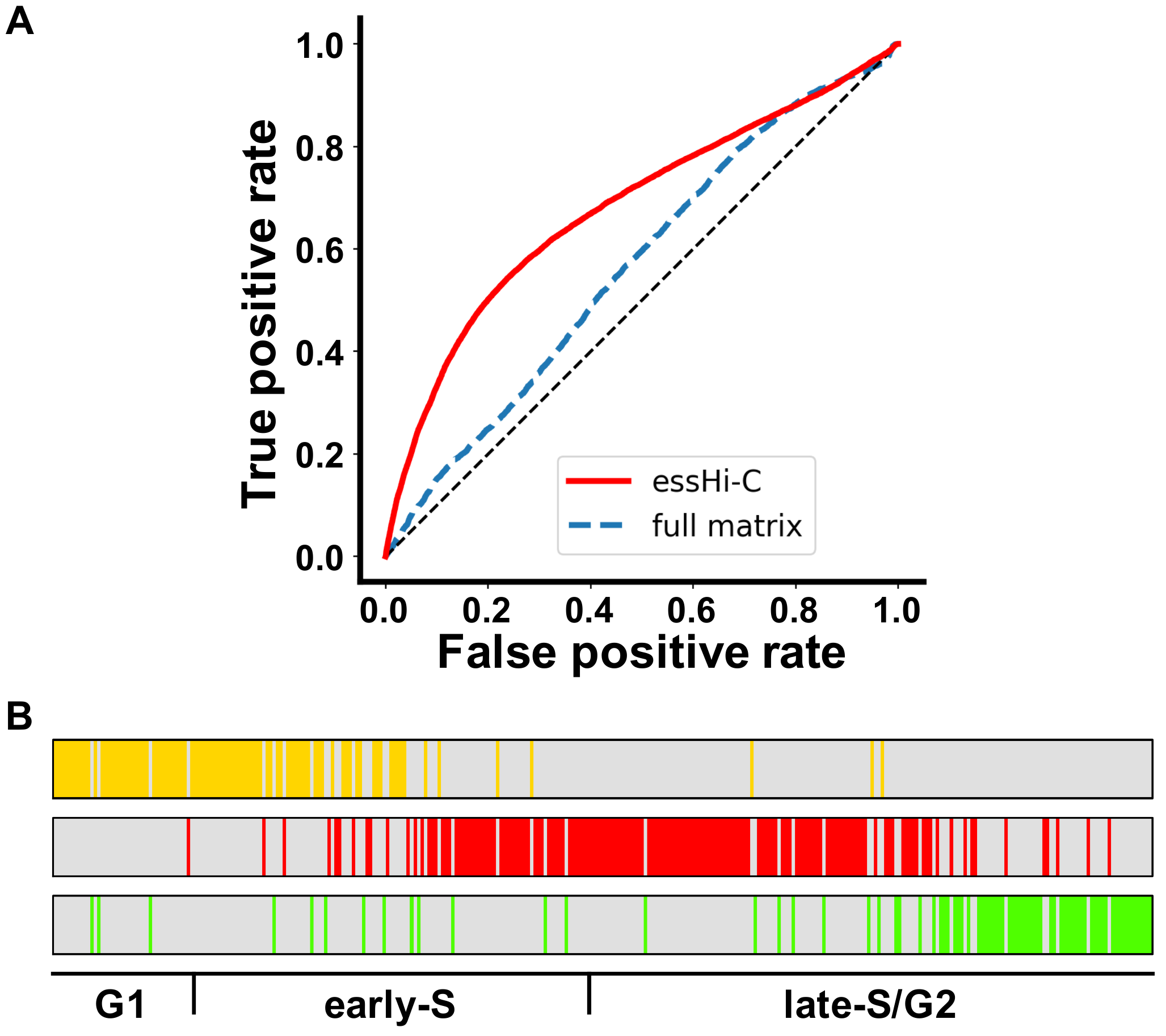}}
\caption{{\bf Cell-phasing of single-cell Hi-C matrices.} {\bf A} The distances of single-cell matrices from \cite{Nagano_2017} in their full and essential forms were used to compute the ROC curves, where true positives correspond to instances with the same G1, early-S, late-S/G2 labelling of \cite{Nagano_2017}.  {\bf B} The bands correspond to the tripartite clustering assignment of the essHi-C matrices. The entries follow the time ordering of \cite{Nagano_2017}.}
\label{Fig4}
\end{figure}

\subsection*{Single-cell Hi-C matrices}

Lastly, we applied the essHi-C analysis to an entirely distinct Hi-C context, namely single-cell Hi-C (scHi-C).
We considered the scHi-C set of ref.~\cite{Nagano_2017}, which covers different cell-cycle stages of the mESC mouse embryonic cell line. In ref.~\cite{Nagano_2017} this data set time-ordered {\em a posteriori} with an elegant  dimensional reduction procedure, which was instrumental to enhance the features of the inherently sparse matrices. To account for the sparsity of scHi-C matrices we performed the essential analysis using $n^*=50$ eigenspaces, see Methods and Supplementary Fig.~S2.

The set of scHi-C matrices, in their plain form, cannot be clustered in a meaningful time-ordered way, as shown by the near-diagonal trend (blue line) in the ROC plot of Fig.~\ref{Fig4} (AUC=0.55).
The essHi-C matrices, instead, show a noticeable and significant correlation, AUC=0.68.
Indeed, the same metrics and clustering procedure adopted for the ensemble Hi-C dataset of Fig.~\ref{Fig3} returns primary partitions that are in very good accord with the time-ordered cellular stages proposed by \cite{Nagano_2017}, see Fig.~\ref{Fig4}B.

\section*{Discussion}
We presented a systematic spectral analysis of Hi-C matrices and established a general strategy to isolate their essential component from the largely-aspecific complementary one.

The starting point of the analysis was the comparison of the spectral properties of OoE-normalised Hi-C matrices with those of random matrices. For the latter we considered the Gaussian orthogonal ensemble, which is commonly taken as reference for the statistical spectral properties of systems with many degrees of freedom with complex interactions.

The comparison presented in Fig.~\ref{Fig1} demonstrated that random matrices are viable terms of references for Hi-C matrices. In fact, the distributions of their eigenvalues and eigenvectors components are largely superposable, except for a small subset of outlier eigenspaces. It is this part of the spectrum, which stands out from the statistically-dominated background, that subsumes the specific features of the Hi-C data. We used to define the essential Hi-C matrix via the spectral summation that is algorithmically implemented in the essHi-C package.

The gist of the essHi-C analysis is different from approaches addressing data noise in Hi-C matrices at the local or bin-wise level. The latter, in fact, are designed to remove biological biases \cite{Yaffe_2011,Vidal_2018} or numerical imbalance \cite{Imakaev_2012,Knight_2012} from Hi-C matrices, while essHi-C discounts the aspecific component by isolating the spectral, hence generally non-local properties that differ from random matrices.

The resulting enhanced specific content of the essential matrices is illustrated by the clearer and sharper features of essential matrices (Fig.~\ref{Fig1} and, especially, by the comparison of different instances of Hi-C matrices from the same or different cell lines (Fig.~\ref{Fig1}G). The subtraction of matrices of the same cell lines is noticeably more uniform and less noisy for the essential matrices compared to the full ones. In addition, the subtraction of the essHi-C matrices of different cell lines provides a neater highlighing of the different features, which are instead convolved with noise in full matrices.

We focussed on two applications of the essHi-C analysis, chosen for their relevance and challenging nature.
Firstly, we compared full Hi-C matrices obtained at high sequencing depth, with matrices at lower depth, both in the full and essential forms, see Fig.~\ref{Fig2}. The comparison demonstrated that essential matrices have a significant boost of correlation with the highest depth reference matrices. In fact, the correlation of the essential matrices is only modestly impacted by the decrease of the sequencing depth. These results provide a striking illustration of the significant potential that the essHi-C analysis for isolating specific interaction features that would require major increase of sequencing depth to be discerned in plain matrices.

Secondly, we carried out the unsupervised clustering of a heterogenous ensemble of Hi-C matrices covering several cell lines, see Fig.~\ref{Fig3}. Good correpondence of cell lines and the unsupervised hierarchical subdivisions are observed only for the essHi-C matrices, not the full ones.
Furthermore, essHi-C based subdivisions of the IMR90 cell lines correlate with the different restriction enzymes used in the Hi-C assays for the two subsets. This unexpected result shows that different experimental probes can reflect in sufficiently distinct contact propensities, and that these can be picked up using essHi-C analysis.

Overall, the results show that essHi-C matrices are better suited than full matrices to isolate significant contact patterns, which ought to be useful also in context where contact propensities are used for chromosome modelling both to generate mean-field genome structures \cite{Trussart_2015,Serra_2015} or to highlight the cell-to-cell variability \cite{Giorgetti_2014,Tjong_2016}.

Finally, to illustrate the perspective potential of essHi-C analysis we discussed a preliminary application to single cell matrices, focussing on the set of ref.~\cite{Nagano_2017}
The ROC curves in Fig.~\ref{Fig4} show that the time ordering of Nagano {\em et al.} cannot be recovered from the full scHi-C matrices. This is consistent with the fact that a dimensional-reduction of scHi-C matrices was needed to obviate to the sparsity of the two-dimensional full matrices and establish their time-ordering. It is therefore significant and appealing that, once the matrices are casted in their essential form, a clear correlation with the time ordering of ref.~\cite{Nagano_2017} emerges, and the main cellular phases are recovered, see Fig.~\ref{Fig4}.
This fact, suggests that essHi-C analysis might be profitably used in place of the dimensional reduction step for a more direct determination of time-ordering or other scHi-C applications.

More in general, our results further emphasizes the advantages offered by essHi-C analysis across very different contexts.

\section*{Conclusion}

We presented a systematic spectral analysis of Hi-C matrices and established a general strategy to
identifying and separating their essential component from the largely aspecific complementary one.

By analysing essential Hi-C matrices in different contexts, we established numerous advantages over the use of spectral-complete (full) ones: from improving the sharpness and clarity of the specific interaction to enhancing the robustness again sequencing depth, allowing the unsupervised clustering of different cell lines and the cell-phasing of single-cell assays.

The results open numerous perspectives for using essHi-C analysis to optimally isolate biologically- and physically-relevant information from Hi-C matrices. Beyond the applications considered here, we expect our tool to be useful in comparative contexts where variations of chromosome compartmentalization could be
picked up with enhanced reliability and hence better related to epigenomics changes \cite{Vilarasa_Soler_2020} or cell differentiation \cite{Bonev_2017,Stadhouders_2018,Paulsen_2019}.

The essHi-C software package is made freely available for academic use and can be accessed at \href{https://github.com/stefanofranzini/essHIC}{https://github.com/stefanofranzini/essHIC}.


\section*{Funding}
This work has been supported by the Italian Ministry for University.\vspace*{-12pt}

\clearpage


\bibliographystyle{unsrt}

\end{document}